# The mind-brain relationship and the perspective of meaning

**Ranjan Mukhopadhyay**

Department of Physics, Clark University, Worcester, MA 01610.

Email: ranjan@clarku.edu

**Abstract:**

We view the mind-body problem in terms of the two interconnected problems of phenomenal consciousness and mental causation, namely, how subjective conscious experience can arise from physical neurological processes and how conscious mental states can causally act upon the physical world. In order to address these problems, I develop here a non-physicalist framework that combines two apparently antithetical views: the materialist view of the mind as a product of the brain and the metaphysical view of consciousness rooted in an underlying hidden reality. I discuss how this framework resolves the problem of mental causation while being simultaneously consistent with fundamental physical principles. I will elucidate how the framework ties in to the perspective of "meaning" that acts as the bridge between physical neurological processes and the conscious mind. Moreover, we will see how both our awareness of the self and our representation of the external world are connected to this perspective.

## 1. Introduction

Despite intense focus on the mind-body problem and despite great progress in cognitive science (see Frankish, 2012), a satisfactory solution to the mind-body problem appears to elude us. Schopenhauer famously called this problem a "Weltknoten," or "world-knot" (Schopenhauer, 1966). From a modern perspective, the reason why it appears insoluble is that rather than being a single problem, it encompasses two interlocking problems: the problem of phenomenal consciousness

and that of mental causation, as highlighted by Jaegwon Kim (Kim, 2005). In his words:
> "The problem of mental causation is to answer this question: How can the mind exert its causal powers in a world that is fundamentally physical? The problem of consciousness is to answer the following question: How can there be such a thing as consciousness in a physical world, a world consisting ultimately of nothing but bits of matter distributed over spacetime behaving in accordance with physical law? As it turns out, the two problems are interconnected—the two knots are intertwined, and this makes it all the more difficult to unsnarl either of them."

Both physicalism and Cartesian dualism have a difficult time unsnarling this intertwined knot. While physicalism appears well poised to explain mental causation, it remains unable to explain how sentience can arise from unconscious physical processes in the brain (this is the hard problem of consciousness, see Chalmers 1995, 1996). While dualism appears to be able to avoid the second problem, it has great difficulty solving the first problem: how can an unphysical mind take input from and causally act on the physical brain. In this paper, we will not discuss Cartesian dualism in any detail but will nevertheless seek an alternative to physicalism. Any such alternative will however have to grapple with our expectation of the causal closure of the physical world. Here we will undertake the task of developing a theoretical framework that can unsnarl the double knot and thus provide a starting point for a theory of consciousness. For this purpose, I will synthesize ideas and research results from a number of disciplines ranging from quantum physics to philosophy to cognitive neuroscience.

To better appreciate the nature of the difficulty, consider the basic neurobiology of the brain. The brain consists of an intricate network of neurons that communicate with each other via a combination of electrical and chemical signals (Bear *et al.*, 2006). It appears reasonable to assume that the brain is carrying out computations in the sense of processing information from the surroundings, the information being transmitted, for example, in the form of electrical impulses, processing this information, and sending out instructions about how the body should react to the environmental cues. Given the immense complexity of the brain, maybe it's not surprising that we do not understand the details of how the brain accomplishes this. The true puzzle is how conscious experience fits into this picture. What rules govern how the brain will interpret certain patterns of neural firings as pain, or pleasure, or sorrow, or joy? How can we explain our sensations in physical terms, given that we do not even understand how to *represent* such subjective mental states in physical or mathematical ways? And yet, if these mental states are non-physical, how can they have any form of causal efficacy without violating physical principles?

Over the years, there have been a number of attempts to deal with this difficult issue (see [Chalmers, 2002], for a systematic treatment of a variety of positions). Rather than attempting a systematic overview, I will briefly outline three currently prevailing philosophical approaches, namely eliminativism, naturalistic dualism, and biological naturalism, in order to highlight the difficulties that we face in constructing a theory of consciousness. Eliminativism, or eliminative materialism, is the radical claim that our ordinary, common-sense understanding of the mind is deeply wrong and that some or all of the mental states posited by common-sense do not actually exist (Churchland, 1984; Ramsey, 2013). While eliminative materialists agree that ordinary mental states cannot be reduced to or be identified with neurological events or processes, the claim is that there is nothing more to the mind than what occurs in the brain. While most discussions regarding eliminativism focus on the status of our notion of belief and other propositional attitudes, some philosophers have endorsed eliminativist claims about the phenomenal or qualitative states of the mind (Dennet, 1978, 1988; Systma, 2009).

Naturalistic dualism, a position that had been advocated by David Chalmers (Chalmers, 1996), accepts that the physical world is causally closed while simultaneously treating qualia as fundamental and irreducible to the physical. In this view, subjective conscious experience exists and is non-physical, but has no causal efficacy. In this sense, subjective conscious experience appears as an epiphenomenon (Jackson, 1982) that does not have the power to directly affect the physical functioning of the brain.

The third position, advocated by philosopher John Searle (Searle, 1998), accepts the reality of phenomenal conscious states, but treats them as emergent systems-level phenomena. The proposal is that, at the neuronal level, the functioning of the brain can be understood purely in physical terms but at a higher systems-level, we have to understand the functioning of the brain in terms of conscious mental states. Consciousness, in this view, is non-physical and irreducible but arises from physical neurological processes.

Arguing for or against phenomenal eliminativism (see Ramsey, 2013, for a discussion) often reduces to the question of whether we have direct access to or direct knowledge of our own subjective states. If we acknowledge that we can have direct knowledge, for example, of our being happy or sad, and that this direct knowledge needs to be accounted for by a theory of consciousness, then it is difficult to consider eliminativism as a viable option. From this acknowledgement, it follows that phenomenal consciousness is not simply a postulate to explain behavior that can be dispensed as a better explanation comes along, as eliminativism seems to suggest, but is rather something that needs to be addressed and presumably explained by a theory of consciousness.

Though not as extreme as eliminativism, the other two positions can also be shown to lead to viewpoints that are counterintuitive. Consider for example our ability to report on our internal subjective states. Both positions indicate that this ability can be explained in physical terms entirely independent of the subjective states themselves. Naturalistic dualism does so by acknowledging the existence of phenomenal consciousness but denying it causal power in the physical world. While biological naturalism acknowledges the causal power of phenomenal consciousness, this causal power is only manifest at the systems level of description. Thus, at a neurological level, we could entirely understand in physical terms our ability or propensity to report on our subjective states, without having to invoke the existence of the subjective states themselves. We could alter this aspect of biological naturalism rather easily by positing that systems-level conscious states could act top-down on the physical neurological states, but this then directly violates the causal closure of the physical world (CCP) and we are back to our first problem. This highlights the challenge that CCP poses for any non-reductive view of consciousness. In this context, I would like to point out that based on our common everyday usage of words such as "cause" and "effect," it might appear reasonable to posit that physical events could simultaneously have both physical and mental causes. However, it is important to bear in mind that the causal framework of classical physics is precise, quantitative, and deterministic, leaving no room for the non-physical to act upon the physical without violating the laws of classical physics. Whether quantum mechanics might offer a way out is an important question, and one to which we shall return in the next section.

The purpose of this discussion was primarily to highlight the difficulties inherent in developing a satisfactory theory of consciousness that is compatible with empirical results. In particular, we find two separate hard problems: for a physicalist framework, the hard problem is that framed by Chalmers about how subjective conscious experience could arise from objective physical processes (Chalmers, 1996). For a non-physicalist framework, the hard problem is to reconcile such a framework with our dual expectation of the causal closure of the physical world and the causal efficacy of the conscious mind (Kim, 2005). In order to overcome these problems and provide a starting point for a theory of consciousness, we will address two questions that we will treat as distinct but clearly inter-related: the first is the more metaphysical question of how phenomenal consciousness can fit in with our concept of physical reality. The second is the more practical question of the relationship between neurological processes in the brain and conscious mental states. I will argue that a deep analysis of quantum mechanics is important for addressing the first question though not directly so much for the second (a quantum theory of the brain is not being proposed here). Moreover, while my answer to the first metaphysical question might seem radically different from Searle's, we will see that the proposed approach towards the second question will be similar to his.

Before proceeding further, I would like to clarify one terminological issue. For this paper, physicalism is defined as the view that all that exists is physical in the sense of being describable, at least in principle, within the framework of the physical sciences, and governed by physical laws/principles. While this might seem unduly restrictive, for all practical purposes it encompasses all fields of knowledge that do not directly relate to consciousness and conscious intentions. For example, leaving aside consciousness and the functional brain, to the best of our knowledge all biological phenomena can also be treated as physical phenomena and in principle can be studied within the mathematical framework of the physical sciences, though the enormous complexity of biological systems can limit the practicality of such an approach. In this paper, in regards to consciousness, I argue against physicalism in this sense of the term. This does not mean the abandonment of naturalism, provided the word is broadly construed. Indeed, the aim of this paper is to provide a starting point for a systematic scientific theory of consciousness that can directly incorporate/integrate empirical results from brain/neurological research as well as psychological research.

## 2. The relation between physical reality and consciousness

As briefly discussed earlier, consciousness does not fit in naturally with our concept of physical reality. In order to understand the relation between consciousness and physical reality, I propose a deeper enquiry into the nature of physical reality, which brings us to modern physics and quantum mechanics. To see why quantum mechanics might have some bearing on the mind-body problem, consider the following. Our usual expectation might be that macroscopic objects (and the brain being an example of such an object) exist objectively and have definite properties (such as position) independent of our minds and our perceptions of such objects. But quantum mechanics calls into question this expectation. Quantum mechanics indicates that we cannot ascribe definite observer-independent attributes, such as position and momentum, to microscopic quantum systems, and the question is whether this extends to macroscopic objects as well. In a 1985 paper, Leggett and Garg introduced an inequality, now known as the Leggett-Garg inequality, to test what they termed as macroscopic realism (Leggett and Garg, 1985). Macroscopic realism, in their view, encompasses two notions: first, that macroscopic objects have definite properties/states independent of observation, and second, that we can measure these states without significantly altering the system. While the experimental tests are technically difficult to perform, recent experiments on mesoscopic (which are nevertheless enormous by atomic length-scales) and macroscopic systems (Palacios-Laloy *et al.*, 2010; Goggin *et al.* 2011, Knee *et al.* 2012, Groen *et al.*, 2013; Zhuo *et al.*, 2015) performed with high precision appear to rule against macroscopic realism. Indeed, a Nature Physics article (Mooji, 2010) describing one such experimental test was entitled "No moon there," referring to Einstein's question of whether we believed the moon existed if no one was observing it (Mermin, 1985). This appears to indicate that even macroscopic systems (such as the physical brain) cannot be said to have objective existence with objective properties independent of the mind. Thus, while discussions of the mind-brain relationship often assume the objective reality of the functional brain, we see that quantum mechanics calls into question this assumption. Thus, the question arises: how should we view this relationship in light of quantum mechanics?

In order to arrive at a reasonable metaphysical framework, we need some *a priori* assumptions that can guide us towards such a framework. I suggest the following:
1. The assumption of an *objective physical reality* that exists independent of us and our perceptions, and that quantum mechanics tells us something about the nature of objective physical reality. If we use the phrase 'empirical reality' to refer to the physical world that we experience, we will not *a priori* assume that that this objective physical reality is identical with empirical reality, but rather let quantum mechanics guide us towards the relationship between the two.

2. The assumption of *phenomenal realism*, implying the existence/reality (at least in some loose sense) of subjective experience. This implies, for example, that for a physicalist, it is a legitimate question to ask how subjective experience can arise from objective physical processes in the brain. This assumption precludes eliminativism and also some versions of reductive physicalism/materialism, and thus our proposed framework should be viewed as an alternative to eliminativism.
3. Any theoretical framework should be self-consistent and consistent with established empirical facts (both from cognitive/brain science and quantum physics).
4. An implicit assumption of simplicity, that is, we seek to find the simplest explanatory framework that is consistent with known experimental results and also consistent with the first three assumptions.

Even with these assumptions, we face the challenge that there is no single interpretation of quantum mechanics universally accepted by physicists. Moreover, the more standard textbook interpretation, known as the Copenhagen interpretation, adopts an instrumentalist approach and avoids entirely these questions about the nature of reality (Faye, 2014). In light of the experiments referenced in an earlier paragraph demonstrating the applicability of quantum mechanics to macroscopic systems, a reasonable starting point would be to adopt what is known as unitary quantum mechanics, which acknowledges the reality of quantum states and assumes that deterministic quantum evolution of states, as described by Schrodinger's equation (or its relativistic counterparts), holds at all scales from the sub-atomic to the scale of the universe. This, for example, is also assumed in much of modern physics, such as quantum cosmology. In a previous paper (Mukhopadhyay, 2014), I had discussed how these assumptions combined with a critical analysis of unitary quantum mechanics lead to a metaphysical framework postulating a deeper level of reality, veiled from direct experience, underlying experienced empirical reality (a similar view was arrived at in Mohrhoff, 2014, from a different starting point). This underlying reality is fundamentally non-local and interconnected. There are two distinct aspects of this reality, the physical that is referenced by quantum physics and the mental that provides the basis for conscious experience. We will use the phrase "underlying proto-consciousness" to reference this second aspect. At the underlying level the two aspects should be treated as distinct, and the mental does not act on the physical, thus preserving CCP at the quantum level. However, at the empirical level the mental and physical are intertwined, so that it is not possible to cleanly demarcate between the objective and subjective. This position could be regarded as a version of dual-aspect monism, in the tradition of the philosopher Baruch Spinoza (see also Atmaspacher , 2014, for modern versions of dual-aspect monism, wherein our position resembles most closely the holistic dual-aspect monism of Pauli and Jung) or as veiled realism: the physicist D'Espagnat's notion of an underlying reality that is veiled from direct experience (D'Espagnat, 2003, 2006). In this view, both physical empirical reality and our individual conscious minds co-emerge (not in a temporal sense) from this underlying level of reality. While we can, for most practical purposes, treat empirical physical reality as objective and observer-independent, the assumption of a consciousness-independent physical reality breaks down when we attempt to understand the relation between phenomenal consciousness and physical brain processes. I note here that the proposed view proposed is entirely distinct from the framework of consciousness-mediated collapse of the wavefunction in the lineage of von Neumann, Wigner, and Stapp (Stapp, 2007).

Following Mukhopadhyay, 2014, I briefly outline the relation between underlying reality and empirical reality. For this purpose, let me briefly introduce the concept of entanglement in quantum mechanics: if two quantum particles interact their states get entangled and remain entangled so that neither has a well-defined quantum state by itself (Horodecki *et al.*, 2009). Entanglement has been demonstrated experimentally by numerous experiments. In this sense, quantum mechanics has an inherent holism (the idea of non-separability of quantum constituents) built into its mathematical structure. If particle A interacts with particle B, their quantum states get entangled; subsequently if B interacts with particle C all three particles are now in one entangled state. In this manner, we expect the entire universe to be essentially in one giant entangled quantum state, described by a universal wavefunction. The idea, then, is that the physical aspect of underlying reality is referenced by the universal wavefunction, and experienced empirical reality corresponds to one branch of this underlying reality. This analysis resembles that of Everett, which is usually associated with the many worlds interpretation of quantum mechanics (Everett, 1957; Barrett, 1999; Wallace, 2012). However, while Everett presupposed physicalism for his analysis, the analysis in Mukhopadhyay, 2014 assumed phenomenal realism, which lead to the different metaphysical view of two levels of reality. At the underlying level, there are no particles in the conventional sense of the word, only quantum fields (Hobbs, 2013). Phenomenal realism then naturally leads to the proposal (Mukhopadhyay, 2014) that, in addition to the physical, there is a second aspect of underlying reality that is fundamentally nonlocal and that provides the basis for our individual conscious minds. This aspect, referenced as an underlying proto-consciousness, does not directly act upon underlying physical reality but plays an important role in the selection of empirical physical reality. While the physical and the mental are distinct at the underlying level, this is not true for empirical reality: instead the analysis suggests interpenetration or intertwining of the objective (physical) and subjective (mental) at the empirical level.

How does the adopted metaphysical framework help us with the double knot of consciousness? In the proposed framework, physical reality is causally closed at the underlying level (thus being compatible with unitary quantum mechanics) but not strictly at the empirical level, thus enabling causal efficacy of consciousness at the empirical level while still maintaining the irreducible nature of conscious experience. The proposed framework simultaneously provides a satisfactory interpretation of quantum mechanics. While this position differs from the materialist principle of causal closure at the empirical level and might thus make us uncomfortable, it is important to bear in mind that it is entirely compatible with all known physical principles. The laws that we directly encounter at the empirical level, such as Newton's laws of motion, should be treated as approximate and statistical laws that emerge from the underlying quantum dynamics. Consider, for example, an analogy with fluid hydrodynamics as governed by Navier-Stokes equation (Batchelor, 1967). This is a deterministic equation and is extremely successful at quantitatively describing a wide range of phenomena associated with the flow of fluids. Nevertheless, even within classical physics this is an emergent law; at the microscopic level the motion of individual molecules would appear quite random. We need to perform a statistical averaging to go from "random" microscopic motion to macroscopic flow governed by the deterministic equation. Under a variety of circumstances, for example, if microscopic randomness gets amplified to a macroscopic level, Navier-Stokes equation could break down without violating any fundamental physical principles. In a similar manner, the framework of classical physics itself should be treated as being approximate and statistical in character, and while the framework might hold with high degree of accuracy for a wide range of phenomena, it could break down in certain circumstances even for macroscopic systems (for example, in the cases of superfluidity and superconductivity, see [Anderson, 1997; Tilley and Tilley, 1990]). I will return to this issue of causal closure in the next section.

## 3. Consciousness and the functioning of the brain

A number of questions could arise in the context of the proposed framework, such as: what are the characteristics of underlying proto-consciousness? How should we understand the precise relationship between this underlying proto-consciousness and our individual conscious minds? While these are natural and fascinating questions, I suggest that from a scientific perspective they may not be the right questions to focus on at this stage. Instead, in order to develop a systematic theory of consciousness tied directly to empirical research, and in order for such a theory to be experimentally testable, it is important that a theory of consciousness be first developed directly at the empirical level. This then brings us back to the second question, which can now be formulated as: at the empirical level, how should we understand the relationship between neurological processes in the brain and subjective conscious states? The proposed metaphysical framework suggests that physical brain processes and conscious mental states should be viewed as intertwined/inter-related but neither is reducible to the other (Mukhopadhyay, 2014). The subjective aspect of conscious experience should be viewed as being fundamental to consciousness and not explainable in terms of the physical. Moreover, since in this framework consciousness is a top-down phenomenon, the unity of experience (often referred to as the binding problem, see Chalmers, 2010, Chapter 10 for an extensive discussion of the unity of consciousness) should also be treated as a fundamental and irreducible aspect of consciousness that cannot be explained in terms of the physical or analyzed further. To quote Chalmers, 2010
"At any given time, a subject has a multiplicity of conscious experiences. A subject might simultaneously have visual experiences of a red book and a given tree, auditory experiences of birds singing, bodily sensations of a faint hunger and a sharp pain in the shoulder, the emotional experience of a certain melancholy, while having a stream of conscious thoughts about the nature of reality. These experiences are distinct from each other: a subject could experience the red book without the singing birds and the singing birds without the red book. But at the same time, the experiences seem to be tied together in a deep way. They seem to be *unified*, by being aspects of a single, encompassing state of consciousness."

It is this phenomenal unity of consciousness (in contrast to what Chalmers calls access unity) that is fundamental in our framework. The notion of the self as distinct from the environment, however, should not be treated as being fundamental in the same manner (unlike in Cartesian dualism), and should be treated as an emergent phenomenon that is analyzable.

Moreover, since in this framework consciousness is a top-down phenomenon, the unity of experience (often referred to as the binding problem, see Chalmers, 2010, Chapter 10 for an extensive discussion of the unity of consciousness) should also be treated as a fundamental and irreducible aspect of consciousness that cannot be explained in terms of the physical or analyzed further. The notion of the self as distinct from the environment, however, should not be treated as being fundamental in the same manner, and should be treated as an emergent phenomenon that is analyzable. We will return to the question of the self in the next section.

What are the implications of our metaphysical view for a systematic investigation of consciousness? Our view lines up quite naturally with the approach of *critical phenomenology* (CP) – a commonsense but nonreductive approach to the study of mind -- proposed by Velmans (Velmans 2006, 2007). While adopting the conventional view that human experiences have causes and correlates in both the external world and the brain, that can be studied by third person methods commonly used in cognitive science, neuroscience, etc., CP recognizes that such methods have to be supplemented by first-person methods that guide subjects to attend to aspects of their conscious experience that are of interest to experimenters. First-person accounts of subjective experience, for example, can inform third person accounts of changes in brain activity, etc. – indeed, such first person accounts would be essential for discovering the neural correlates of said experience.

In order to relate the proposed framework to brain science, and given the remarkable advances over the past few decades in our understanding of neurobiology and cognitive neuroscience, it makes sense to take a more neuro-centric approach, which is what we will adopt for the remainder of the paper. It is important to bear in mind that in our framework consciousness does have causal efficacy; the hypothesis is that if a system is conscious, this *will* be manifested in the system's behavior so that we will not be able to analyze or understand its behavior in purely physical terms. So, the question that arises is: how does the non-physical nature of consciousness manifest itself in the physical activities of the brain? Before addressing this question, let us briefly return to the issue of causal closure of the physical. In a somewhat imprecise way we could think of causal closure in the following way: at the empirical level the physical is not causally closed at the level of individual atomic scale micro-processes, the laws of classical physics emerge by averaging over a large number of such micro-processes. For a wide range of macro-systems, classical physics applies with remarkable accuracy and we can assume for all practical purposes the causal closure of classical physical reality. However, in situations where coherence develops between such micro-processes, as in superfluidity and superconductivity, the classical laws could break down even at the macro-scale (Anderson, 1997; Tilley and Tilley, 1990). My hypothesis is that for consciousness to have causal efficacy in any significant manner, there similarly has to be a level of coherence that has to develop in the neural activities of the brain (neural coherence); however, unlike superconductivity, the origin of this coherence cannot be understood purely in physical (quantum) terms. Such a hypothesis is indeed consistent with the observation that conscious activity is correlated with temporal coherence in neural firing across a large population of distributed neurons (Singer, 2007). Indeed, this could be treated as a prediction of our model: for subjective consciousness to have causal efficacy, long-range coherence in neural activities is a necessity. While the connection between consciousness and coherent neural activity is not new and has been proposed, for example, in the context of the global workspace theory (Dehaene, 2011) (though, within a physicalist framework, we are still left with the question about the mechanism by which the brain interprets neural coherence in terms of conscious experience), in our framework such neural coherence is essential for causal efficacy of consciousness but not necessarily for conscious experience itself. While the two typically go together, there might be cases involving patients under anesthesia, or brain trauma, or near death experience, where it might be possible to have conscious experience without causal efficacy.

Let us then postulate that consciousness should be associated with physical activities of the brain at the systems level (and not at the level of individual neuronal activities). This appears similar to biological naturalism posited by Searle, with one important difference. The difference is that in our framework the systems-level dynamics supervenes on neuronal activities in a manner that makes it impossible to understand the functioning of the brain completely in physical terms both at the systems level and also at the neuronal level. Since this point can cause some confusion, let me explain the proposed viewpoint by introducing two contrasting examples. As our first example, consider the collective behavior of an ant colony (Hölldobler and Wilson, 1990). Even though each ant performs its own simple actions and follows relatively simple local rules, the colony as a whole can build astonishingly complex structures that are of great importance for the survival of the colony as a whole. Clearly the behavior of the colony cannot simply be reduced to the behavior of the individual ants. Nevertheless, to borrow some terminology from David Chalmers (Chalmers, 1996), we expect reductive explanations to still work in the sense that if we understand precisely the rules that each ant follows, in principle we can model precisely the dynamics of the colony as a whole. Thus, we expect a theoretical or computational model to exist that would explain the behavior of the colony in terms of the dynamical rules of individual ants (factoring in, of course, their mutual interactions). But now, by contrast, consider the evolutionary dynamics of human society. A little thought should clarify that such a reductive explanation may not work in this case. Because the behavior of human individuals is strongly influenced by systems level properties such as cultural norms and values, we cannot even formulate dynamical rules for the behavior of the individual independent of society as a whole. Thus, while simple reductionism fails for both examples, we expect reductive explanations to work for the first example but not the second. My postulate is that were we to attempt relating systems-level behavior of the human brain to the dynamics of individual neurons, we face a situation akin to the second example, so that reductive explanations of brain functioning also breaks down.

Let us analyze this in some more detail. Borrowing from Fritjof Capra (Capra, 2002), we can understand a system in terms of three interconnected perspectives that he calls structure/matter, form, and dynamics/process. Structure refers to the constituents of the systems and their properties, form refers to the network of relations between the constituents, and dynamics obviously refers to the dynamics of the system as a whole. These three perspectives are important for analyzing any complex system, living or non-living; but the claim is that for analyzing the functioning of the human brain, we need an additional fourth perspective. We can easily understand this fourth perspective by referring back to the second example of human society. If we want to understand the dynamics of a group of humans, it becomes clear what is missing, it is what Capra characterizes as the perspective of "*meaning*." Consider a case of conflict, where entirely different meanings are attached to the same set of events. Or consider sports, where without this fourth perspective, the behavior of a sportsperson may make no sense at all. I propose this is precisely the missing perspective for brain activity as well, but with one important difference. In the case of social dynamics, "*meaning*" is generated at the level of the individual even though how it is generated depends strongly on society as a whole. In the case of brain functioning, however, meaning is entirely generated at the systems level and not at the level of individual neurons. Thus, the task of understanding brain functioning becomes the task of understanding how to integrate the fourth perspective of meaning with the other three perspectives. And the crucial point is that while the other three perspectives can be understood in physical terms, the fourth perspective is non-physical in that it cannot be treated within the abstract mathematical framework of physics, which thus has nothing directly to say about *meaning*. I suggest that it is entirely reasonable to assume that, as an additional perspective, *meaning* arises only in the context of consciousness and plays no direct role in the behavior of non-conscious physical systems that form the subject matter of physics. While *meaning* is occasionally invoked in other contexts, for example, in the context of the immune system distinguishing between "self" and "nonself" cells (Medzhitov, 2002), in those cases meaning is not a necessary and additional perspective. In the case of the immune system, for example, one can understand its functioning in terms of physical mechanism without the need to invoke *meaning*, while I suggest that is not the case for the human brain.

It is important to contrast the proposed theoretical framework with Cartesian dualism; our framework does not imply a non-physical entity residing in the brain guiding the firing of individual neurons any more than the nonreductive nature of human culture implies the existence of a non-physical entity present in human society guiding the action of individuals. Nevertheless, we cannot understand the behavior either of the conscious brain or of human society purely in physical terms in the sense that we defined "physical" in the Introduction. And it is this fourth non-physical perspective of *meaning* that leads to the breakdown of reductive explanations both for the conscious brain as well as for human society. Also, we note here that while the other three perspectives can be directly queried/observed in the laboratory, the perspective of meaning cannot directly be observed or measured. *Structure*, for example, can be studied by examining the behavior of single neurons in the laboratory both individually as well as within small neural networks. The human connectome project would study neural interconnections/wiring in the brain (*form*) and its relationship to function and dynamics. Techniques such as fMRI can study *process/dynamics* by tracking blood flow patterns in the brain (Huettel *et al.*, 2009). The perspective of *meaning*, however, cannot be tracked directly and has to be inferred from behavior/reporting of the subject, matched with observed brain activity, neural connectivity/wiring etc., and combined also with our shared human perspective (we directly understand, for example, what it feels like to be in pain or in a happy state).

Even though we cannot quantify or measure "*meaning*" directly, a moment's reflection should convince us that how we react to events in our lives depend strongly on the meaning that we attach to the events. The same event could generate very different responses depending on how it is interpreted. At an even more basic level, the way we perceive, for example, the world around us has to be dependent on how the brain ascribes meaning to certain neural activities. We seem to be directly aware of objects outside our "selves" but it is clear that the brain can directly only have access to internal neurological signals; it has no access to the external world independent of that. So, at the most basic level, when we are aware of an external object, that awareness has to emerge from how the brain interprets or ascribes meaning to some set of neurological activities. Thus, our perception of the external world, be it visual or auditory or tactile, has to be dependent on how meaning is ascribed to neural activities. Similarly meaning also has to be ascribed to other neurological states that are tied to our internal mental states (for example, states of pleasure or pain, happiness or sorrow). Consider the recent discovery that some of the same set of neurons fire when we are in pain as well as when we observe someone else in pain, yet our conscious experience is quite different: in one case, we directly feel pain whereas in the second case we feel empathy for the other person in pain (Singer, 2004; Lamm *et al.*, 2007, 2010; Jackson, 2005). Very different meanings can be ascribed to similar sets of neural firings.

At the most basic level, the claim then is that the perspective of *meaning* would be very important for our understanding of the neural code: that is, how neural spikes encode information about the external world and also about our internal states. This is in line with current research in computational neuroscience where this perspective is often couched within the notion of the homunculus (little man located in the brain). Quoting from the book by Princeton biophysicist, William Bialek, and collaborators (Rieke *et al.*, 1999):

> "On the other hand, as explorers of the nervous system we place ourselves, inevitably, in the position of the homunculus – we observe the responses of sensory neurons and try to decide what these responses could mean to the organism. This problem of assigning *meaning* [emphasis added] to the activity of sensory neurons is the central issue in our discussion of the neural code."

This is consistent with our proposed framework provided that the notion of a separate homunculus is not taken too literally.

Then, in addition, there is a second layer of meanings related to how we interpret or ascribe meaning to the perceived external world, to language, to our relation with others, etc. Consider, for example, the case of Capgras delusion where the afflicted person says that a friend, spouse, parent, or other close family member (or pet) has been replaced by an identical-looking impostor (Ramachandran, 1998, 2011; Elise *et al.,* 2001).  The direct perception of that person or pet has not changed, but the meaning associated with that perceived person or pet has changed. Analysis of this syndrome (Ramachandran, 2011) indicates that this second level of meaning closely ties our emotional states with our perception of external objects or people.  Our symbolic representation of the world, as well as language semantics are also tied to this second/higher level of meaning.  It is also central to our attempts at understanding the functioning of mirror neurons and the controversy surrounding their role in helping us understand the actions and intentions of other people. Finally, a word of caution. In our framework, the perspective of *meaning* acts as the "bridge" connecting physical brain activity to conscious experience. This does not however imply that conscious experience can be reduced to this perspective. In the proposed framework, the quality of subjective experience should still be treated as fundamental and irreducible, though dependent upon physical brain activity.

While, according to the usage adopted in this paper, the perspective of *meaning* should be treated as non-physical, this does not preclude a systematic and naturalistic theoretical framework that could successfully integrate *meaning* with the other three perspectives. One suggestion might be to adopt formalism from semiotics or semiology – the study of meaning-making, the study of sign processes and meaningful communication – developed by Saussere and others (see Chandler, 2007, for a review). Theoretical research in that direction has already been undertaken by researchers such as Zlatev, Thompson, and others (Zlatev, 2009; Thompson, 2010), giving rise to the emerging field of cognitive semiotics (CS).

The proposed framework can be viewed as a synthesis of two apparently antithetical positions. The first is the view from materialism of treating the conscious mind as a product of the brain, which can be still held to be valid for most practical purposes as long as we are cognizant that the functioning of the brain cannot be understood in purely physical terms. (In this respect, my use of the word physicalism could differ from the way it has been used sometimes by philosophers of the mind, since by our definition, and somewhat non-intuitively, the discipline of neurobiology does not fall entirely in the physical domain, which should be viewed as a consequence of the interpenetration of the subjective and objective.) The second is the ancient mystical/philosophical view of consciousness being rooted in an underlying hidden reality (for example, in Indian Vedantic philosophy, see Easwaran, 2007). While still true in our framework, it is incomplete in the sense that it leaves unanswered the question of how we should understand the relationship between physical brain processes and conscious mental states.

## 4. Perception and the emergent self

The notion that I will explore in this section is that our sense of the self and our awareness of the external world are closely tied with each other, and strongly connected to the perspective of meaning. First let us analyze briefly the notion of the self. Consider the following: First we know that in some sense both the external world as we experience it and our internal world are constructed from our brain activity. Consider the case of dreams, which are all internally generated, and yet we do not experience dreams as all taking place within us. Even in our dreams we have a sense of an external world and the self who experiences this external world. Secondly, we know both from reports throughout human history as well as from recent research in neurotheology that in meditative states, people can entirely lose their sense of self (d'Aquili and Andrew B. Newberg term this as the state of Absolute Unitary Being, d'Aquili *et al.*, 1998; Newberg *et al.*, 2002), highlighting the distinction between unity of consciousness, which remains intact, and the notion of self. Finally, consider also the case of alienated self-consciousness often associated with schizophrenia, where people experience their own conscious episodes as those of another person or agent (Hoffman, 1986; Stephens, 2007). For example, a patient might claim that he has thoughts but they are not his thoughts, the thoughts belong to some third party. The question is how can such a statement even be cogent. How can the thought exist in the patient and the patient be completely aware of that, and yet insist it is not his thought? Hoffman (1986) argues that thoughts, as inner speech, seem alien because the patient experiences them as personally unintended. That is, the patient does not have the sense that he intended to say to himself what the voice tells him. If we are to make sense of what the patient means by statements such as this, the suggestion is that a thought may be mine in two different senses. One consists in experiencing myself as subject of the thought, that is, the thought is present in me. The second way consists of myself as actively involved in the thought, as the thinker or agent behind. This brings us to an important and fundamental aspect related to our notion of the self: the self as the *agency* for our actions. It is this sense of self that gets affected in this disorder, and thus the patient is not saying something incoherent or inconsistent. The suggestion then is that our notion of our self is intricately tied to this second sense of the self as an agent, and the thesis we would like to explore is that this sense of self is linked with our perception of the external world.

Let us turn now to the issue of perception. There has been a growing awareness that perception is a far more active process than might have been suspected earlier (see, for example, Bajczy, 1998; Noe, 2005; Thompson, 2010; Lauwereyns, 2012). While the issue of active perception and its relation to internal representation has been controversial, it is clear that perception requires a very active gaze (see Lauwereyns, 2012, for a fascinating discussion). Without going into a thorough discussion of the intricacies of the surrounding issues, I will adopt here a position somewhat similar to that proposed in Lauwereyns, 2012, that combines classical computational theories of perception (Marr, 1981) with accounts emphasizing the pervasive sensorimotor nature of perceptual experience (Noe, 2005). To understand the active nature of perceptual representation, let me use a thought experiment to develop an analogy with tactile representation. Imagine a person having recently lost the use of his eyesight and being placed in unfamiliar new surroundings that he is to now call his home. In the beginning, he will be completely lost and will keep tripping over or bumping into everything, but over time, through movement, and groping, and grasping, will build up a tactile representation or model of his new home. It will be a predictive model. He might, for example, first fix a location and a way to orient himself at this location. Then his representation may inform him, for example, that if he were to walk in some particular direction, he will bump into some large object. If around that point, he was to turn left he will encounter something soft on which he could sit or lie down. And so on. Thus, it will be a predictive representation for his experience which will inform him that if he were to act in manner A, he will experience B. Notice that central to this representation is the assumption of the stable self as an agent acting on the surrounding and also the assumption of a relatively stable environment that exists independent of the self and his actions. The stability of the self and surrounding is maintained in this representation despite the continuously changing tactile experience.

I propose that our visual perception of the external world works along similar lines. We know already that rapid (saccadic) eye movement is an integral component of human vision, with around 2-4 saccadic movements per second. The idea then is that the brain creates a predictive model for our visual experience which predicts that as we rapidly move our eyes and shift our gaze around some object in manner A, we will receive this series of sensory inputs in manner B. And central to this representation is both the notion of the self as *agent* and also the environment as existing independent of the self. The stability of the environment is represented in terms of objects (as graspable units of visual perception). Thus, for visual perception, the brain has to distinguish between neural motor outputs that are interpreted as coming from the 'self' and sensory inputs that are interpreted as arriving from the external world via our sense organs, and both our notion of a stable self and an independently existing external world are tied to the dynamics between input and output signals. This hypothesis is consistent with the high degree of feedback pathways between all stages of visual processing regions in the brain (Ramachandran, 1998, 2011). The hypothesis also helps explains visual stability across saccades. We know empirically (Strasburger *et al.*, 2012) that at any instant, our visual field consists of two regions: a small central section with high visual acuity and image resolution (foveal vision) and the surrounding area with far reduced visual acuity (peripheral vision). Yet, despite the almost continuous, rapid and dramatic changes in the foveal visual field accompanying the saccades, our perceptual experience remains stable and we have no cognizance of these rapid changes (see Moronnel and Burr, 2008; Lauwereyns, 2012, for more detailed discussions of visual stability). It is difficult to reconcile the stability of vision with any passive theory of perception but follows quite naturally from the model of active perception proposed above (recall that in the thought experiment, the tactile representation remains stable despite the continuously changing nature of tactile sensory input).

I should also point out here a crucial difference between tactile representation in our thought experiment and visual representation: in the tactile case, what the person was aware of were the raw tactile sensations, and the construction of the tactile representation was a slow and largely conscious process. For regular visual perception, we are aware not of the raw sensory inputs but the processed sensations that are tied to the model developed, thus explaining, for example, the persistence of a number of optical illusions even upon gaining conscious knowledge of the nature of the illusion. The process of the development of a predictive visual model itself is rapid, automatic, largely unconscious, and, I propose, underlies both our notion of the self and the notion of the external world as existing independent of the self. While both these notions are also in a sense *a priori* for our visual experience, this does not mean that these are necessarily biologically programmed in us. Instead the hypothesis is that these notions emerge in the course of the infant's development (and maybe the process starts even before birth, at the fetal stage in the mother's womb) as a result of his or her interactions with the environment coupled to the infant's (at least in part biologically programmed) brain development.

At this point, let us return to the question: How does the sense of self tie in to the perspective of *meaning*? It is clear that the brain has to attach meaning to certain nerve impulses as arising from the external world and others as arising from the self. This is true also during dreams and hallucinations where, even though all the nerve impulses arise from within the brain/neural system, certain impulses are still interpreted as being generated from the external world. The cases of alienated self-consciousness clearly suggest that the manner in which meaning is attached to such impulses can be disrupted in the case of mental disorders. Leaving aside, however, such pathological cases, in general which nerve impulses are to be interpreted as input signals and which are to be output signals cannot be understood by looking at the structures and properties of the neurons themselves, instead we have to understand that in terms of how the brain is wired to the sense organs and thus connected to the environment. Based on these observations, it is natural to hypothesize that both how meaning and how the sense of a stable self are generated in the brain cannot be understood by studying the brain by itself in isolation but rather in terms of the relationship between the brain (and more generally the body) and the external world. And due to the close relation between *meaning* and the "self," it is also natural to expect two layers or levels of self, the basic physical self and the social self. The basic self is connected to perception and our relationship with the physical environment while the social self, tied to the postulated second layer of meanings, is connected to our relationship with the social environment. The social "self" references how we view ourselves as social beings, with language, moral values, talents etc., but this view arises from how we relate to others in the social environment: our family, close kin, friends, professional colleagues, and so on (see, for example, Musholt, 2015). And, while the two selves are obviously not disjoint, the proposal is that it is helpful to think of the social self as building upon our basic notion of the self as an agent enduring over time. I should also add here a note of caution: this should not be taken to imply that developmentally these two selves emerge sequentially: the basic self, followed by the social self. Instead their development might even start almost simultaneously, though the development of the social self is far more drawn out and lasts through childhood and presumably into adolescence. A full discussion of the social self in all its complexity and its emergence lies beyond the scope of this paper.

## 5. CONCLUSIONS

The mind-body problem has remained one of the central problems in philosophy. As discussed in the introduction, both physicalist and non-physicalist frameworks have had a difficult time resolving the intertwined problems pertaining to the emergence of subjective consciousness and to the causal efficacy of conscious mental states. In this paper, I have developed a framework which can successfully unsnarl the knot and provide a convenient starting-point for developing a systematic theory of the mind and consciousness, thus converting a seemingly unsolvable philosophical mystery to a problem that can then be addressed using the methods of neurobiology, psychology, and cognitive neuroscience. What might appear somewhat unusual about the proposed framework is the simultaneous acknowledgement (a) of the reality and irreducibility of subjective experience, and (b) of the causal efficacy of subjective consciousness, without violating established fundamental laws of physics or physical principles. While non-physicalist, the proposed framework is compatible with scientific naturalism. Within our framework, we have discussed how the perspective of "meaning" can be the bridge connecting physical processes in the brain to the conscious mind. Finally, we have also discussed the implications of our framework for understanding the relation between our notion of the self and that of an independently existing external reality.